**Spectral behavior of localized plasmon resonances in the near- and far-field regimes. Comment on "*The spectral shift between near- and far-field resonances of optical nano-antennas*",**

F. Moreno,[*,†] P. Albella,[‡] and M. Nieto-Vesperinas[§]

[†]*Grupo de Óptica, Departamento de Física Aplicada, Universidad de Cantabria, Avenida de Los Castros s/n, 39005 Santander, Spain*
[‡]*Experimental Solid State Group, Department of Physics, Imperial College London, London SW7 2AZ, United Kingdom*
[§]*Instituto de Ciencia de Materiales de Madrid, Consejo Superior de Investigaciones Científicas, CSIC, Campus de Cantoblanco, 28049 Madrid, Spain*

*Corresponding author: morenof@unican.es

### Abstract

In a recent paper by Menzel et al. (Opt. Exp. **22**, 9971 (2014)), its authors analyze the spectral red-shift effect of plasmonic resonances of metallic nano-antennas between the far-field and near-field regimes. Here, we demonstrate that their interpretation of this effect is done under the same perspective as one recently reported in Langmuir **29**, 6715 (2013); however, the former is incomplete and needs some remarks and clarifications which require additional relevant concepts and arguments of physical significance.

In a recent paper [1], C. Menzel et al. analyze the spectral red-shift of the plasmonic resonances in metallic nano-antennas between the far-field and near-field regimes. This interesting phenomenon is analyzed under a hypothesis which considers both a spectral Lorentzian-lineshape of the dipole moment amplitude $|\vec{p}(\omega)|$ and the near field emission spectrum according the well-known expression for its electric vector at an arbitrary point defined by the unit vector $\vec{n}$

$$\vec{E}_s(\vec{r}) = \frac{1}{4\pi\varepsilon_0}\left[\vec{p}\left(k^2 - \frac{1}{r^2} + \frac{ik}{r}\right) + \vec{n}(\vec{n}\vec{p})\left(-k^2 + \frac{3}{r^2} - \frac{3ik}{r}\right)\right]\frac{e^{ikr}}{r}. \qquad (1)$$

According to these authors, the far-field contribution can be easily identified, scaling as $\vec{p}k^2$ and hence the corresponding intensity as $|\vec{p}|^2 k^4$. From this consideration, they conclude that the maxima in the near-field intensity can be obtained from the measured scattering cross section by just dividing it by $\omega^4$.

This simple interpretation needs some remarks and clarifications, and requires additional concepts and arguments of physical significance [2].

Although the authors of [1] seemed initially unaware of the analysis of [2], and they later kindly quoted ref[2] in their page proof process, pointing out their different approach, it should be emphasized that in fact [1] employs the same perspective as [2], even including the additional discussion of [1] for Mie spheres or cylinders. The purpose of this comment is to clarify this, so that future researchers are not confronted and misled by the existence of two apparently different theories [1, 2] that are, as a matter of fact, the same. In this sense, we show in the following that the analysis of [1] is a part of the theory and concepts developed in [2].

Ref [2] starts with the representation of the scattered electric vector by a spherical particle in terms of plane wave components by means of the angular spectra $S_1(\cos\alpha)$ and $S_2(\cos\alpha)$ for the two possible polarizations: normal and contained in the scattering plane defined by the unit vectors $\vec{e}_\alpha$ and $\vec{e}_\beta$, respectively and according to the equation

$$\vec{E}_s(\vec{r}) = \frac{1}{2\pi}\int_0^{2\pi} d\beta \int_0^{\frac{\pi}{2}-i\infty} e^{i\vec{k}\vec{r}}\left[S_1(\cos\alpha)\cos\beta\,\vec{e}_\alpha - S_2(\cos\alpha)\sin\beta\,\vec{e}_\beta\right]\sin\alpha\,d\alpha. \qquad (2)$$

For a spherical non-magnetic dipolar particle in the wide sense, (i.e. that whose scattering may be fully described by its first electric Mie coefficient $a_1$), Eq.(2) reduces to (see [2])

$$\vec{E}_s(\vec{r}) = \frac{3a_1}{4\pi}\int_0^{2\pi} d\beta \int_0^{\frac{\pi}{2}-i\infty} e^{i\vec{k}\vec{r}}\left[\cos\beta\,\vec{e}_\alpha - \cos\alpha\sin\beta\,\vec{e}_\beta\right]\sin\alpha\,d\alpha \qquad (3)$$

We must draw the attention of readers to the fact that the electric field vector, as written in either Eq.(2) or Eq.(3), with which all calculations are done in [2], has been chosen such that it has no dimensions, in contrast with that usually employed, (cf. Eq. (1) above, referred as Eq. (2) in [1], which goes as $r^{-2}$, and Eqs. (6)-(8) of [2]). This can be immediately seen since $a_1$ in Eq.(2) and Eq.(3) is dimensionless, and thus the scattered electric field (3) may be written in the form (1) but with $\vec{p}(\omega) = \varepsilon_0 \alpha'_E \vec{E}_0$, where $\vec{E}_0$ is the field incident on the particle whose polarizability is $\alpha'_E = 6\pi i a_1$. Notice this latter dimensionless expression of $\alpha'_E$ instead of the usual polarizability $\alpha_E$ which has dimensions of volume, $\alpha_E = 6\pi i k^{-3} a_1 = k^{-3} \alpha'_E$.

As a matter of fact, in this notation Eq. (3) yields the far field (*FF*) given by the asymptotic value for large *kr* values of the homogeneous part of the integral (3), (cf. [2])

$$\vec{E}_{FF} = k^{-1}(\vec{n} \times \vec{p}) \times \vec{n} \frac{e^{ikr}}{r}. \tag{4}$$

$\vec{E}_{FF}$, as given by Eq. (4), is $k^{-3}$ times the usual expression for the far field derived from Eq. (1) and employed in [2]. Consequently, as pointed out in [2], in connection with the calculations shown in its Fig. 4, the *FF* intensity $|\vec{E}_{FF}|^2$ scales as $k^{-2}$, or equivalently as $\lambda^2$. Correspondingly, the near field (*NF*) $\vec{E}_{NF}$ of the dipolar particle is given by the full integral (3) containing both propagating and evanescent components (cf. [2]), and hence it may be expressed as

$$\vec{E}_{NF} = i \int_0^{2\pi} d\beta \int_0^{\frac{\pi}{2}-i\infty} [(\cos\alpha)^{-1}(\vec{n} \times \vec{p}) \times \vec{n}] e^{ikr} \sin\alpha \, d\alpha. \tag{5}$$

But as it was pointed out in Fig. 4 of ref [2], the intensity of the near field (5) scales as $\lambda^6$, i.e. as $k^{-6}$; namely, $\vec{E}_{NF}$ scales as $k^{-3}$; (this dependence with $\lambda^6$, makes the plasmon resonance to be red shifted in the near-field regime, and consequently makes this shift weaker or stronger depending of the spectral characteristics of the resonance according to Eq.(3) (or equivalently to Eq.(4) of [2]). Thus, the difference of power in *k* between the far field intensity $|\vec{E}_{FF}|^2$ that scales as $k^{-2}$ and that of the near field $|\vec{E}_{NF}|^2$ that scales as $k^{-6}$, is a factor $k^{-4}$, which is exactly the same conclusion as that obtained in ref [1].

Bearing in mind all previous considerations, three additional remarks should be made. First, the above $k^{-4}$ factor between near field and far field region intensities is underlined by the contribution of evanescent waves to the former wave-field. These modes are physical entities essential to characterize such fields. Therefore, they are at the root of that difference of *k*-scaling of the field between these two regions.

Second, as a consequence of the latter point, these evanescent components, which dominate in the near field, yield in this region the $k^{-4}$ factor in the intensity not only in the electrostatic limit $\lambda \to \infty$, namely $|Im(\alpha_E)| \gg 0$, [cf. Eq. (2) above], which is the one where the analysis in [1] is confined, (only the limiting electrostatic approximation is addressed in [1] to describe the near field), but also all along the imaginary axis, $\frac{\pi}{2} - iIm(\alpha_E)$ of Eq. (2), (i.e. in any region of the near field beyond that described by electrostatics), in which the contribution of evanescent components dominates over that of the propagating ones.

Finally, since as stated above the particle dipole moment $\vec{p}(\omega)$ is expressed as $\vec{p}(\omega) = \varepsilon_0 \alpha'_E(\omega) \vec{E}_0$, with $\alpha'_E(\omega)$ described by the Mie coefficient $a_1(\omega)$ as $\alpha'_E(\omega) = 6\pi i a_1(\omega)$ and therefore it has a frequency dependence given according to the damped peak described by the Mie resonance of $a_1(\omega)$ (no other extra assumption is necessary [3]), we conclude that a phenomenological description of the dipole excitation like that of a harmonic oscillator, as given by Eq. (1) in [1], unnecessarily loses rigor and accuracy.

**Acknowledgments**

MN-V acknowledges support from the Spanish Ministerio de Economía y Competitividad (MINECO) through FIS2012-36113-C03-03 research grant. FM acknowledges support from USAITCA (award #W911NF-13-1-0245). PA acknowledges the support received from the Leverhulme Trust.

**References**

[1] C. Menzel, E. Hebstreit, S. Mühlig, C. Rockstuhl, S. Burger, F. Lederer, and T. Pertsch, "The spectral shift between near- and far-field resonances of optical nano-antennas" . *Optics Express*, **22**, 9971–9982 (2014)

[2] F. Moreno, P. Albella and M. Nieto-Vesperinas, "Analysis of the spectral behavior of localized plasmon resonances in the near- and far-field regimes", *Langmuir* **29**, 6715–21 (2013)

[3] J.M. Geffrin, B. García-Cámara, R. Gómez-Medina, P. Albella, L.S. Froufe-Pérez, C. Eyraud, A. Litman, R. Vaillon, F. González, M. Nieto-Vesperinas, J.J. Sáenz, and F. Moreno, , "Magnetic and electric coherence in forward- and back-scattered electromagnetic waves by a single dielectric sub-wavelength sphere", *Nature Communications*, **3**, 1171 (2012). doi:10.1038/ncomms2167